\documentclass[journal=jctcce,manuscript=article,layout=twocolumn]{achemso}
\setkeys{acs}{articletitle=true}

\usepackage[T1]{fontenc}
\usepackage{amsmath, amssymb}
\usepackage{siunitx}
\usepackage{xcolor}
\usepackage{braket}
\usepackage[version = 3]{mhchem}
\usepackage{paralist}
\usepackage[normalem]{ulem}

\newcommand{\onlinecite}[1]{\hspace{-1 ex} \nocite{#1}\citenum{#1}}

\let\oldmaketitle\maketitle
\let\maketitle\relax

\title{Second-order exchange-dispersion energy based on multireference description of monomers}
\author{Micha{\l} Hapka}
\email{hapka@tiger.chem.uw.edu.pl}
\affiliation{Institute of Physics, Lodz University of Technology, ul. Wolczanska 219, 90-924 Lodz, Poland}
\alsoaffiliation{Faculty of Chemistry, University of Warsaw, ul. L. Pasteura 1, 02-093 Warsaw, Poland}
\author{Micha{\l} Przybytek}
\affiliation{Faculty of Chemistry, University of Warsaw, ul. L. Pasteura 1, 02-093 Warsaw, Poland}
\author{Katarzyna Pernal}
\affiliation{Institute of Physics, Lodz University of Technology, ul. Wolczanska 219, 90-924 Lodz, Poland}

\begin{document}


\twocolumn[
\begin{@twocolumnfalse}
\oldmaketitle
\begin{abstract}
We present a method for calculation of the second-order exchange-dispersion energy in the framework of the symmetry-adapted perturbation theory (SAPT) for weakly interacting monomers described with multiconfigurational wave functions. The proposed formalism is based on response properties obtained from extended random phase approximation (ERPA) equations and assumes the single-exchange ($S^2$) approximation. The approach is applicable to closed shell systems where static correlation cannot be neglected or to systems in nondegenerate excited states. We examine the new method in combination with either generalized valence bond perfect pairing (GVB) or complete active space self consistent field (CASSCF) description of the interacting monomers. For model multireference dimers in ground-states (\ce{H2}$\cdots$\ce{H2}, Be$\cdots$Be, \ce{He}$\cdots$\ce{H2}) exchange-dispersion energies are reproduced accurately. For the interaction between the excited hydrogen molecule and the helium atom we found unacceptably large errors which is attributed to the neglect of diagonal double excitations in the employed approximation to the linear response function.
\end{abstract}
\end{@twocolumnfalse}
]


\section{Introduction}

Decomposition of the intermolecular interaction energy into physically meaningful contributions requires partitioning of the system into fragments, the mutual interaction of which may be described either with variational of perturbational methods.\cite{Stone:97,Gordon:12} In symmetry-adapted perturbation theory (SAPT)\cite{Jeziorski:94,Szalewicz:05} the interaction energy is expanded in the intermolecular interaction operator and the proper permutational symmetry of the total wave function is imposed by the use of antisymmetrizer operators in each order. Symmetry forcing leads to the emergence of exchange energy corrections which are vital for the correct description of the short-range repulsion.

In contrast to the supermolecular method, in SAPT one computes the interaction energy directly, based solely on monomer properties. Since the dimer calculation is never performed, one avoids the basis set superposition error. Over the last four decades, SAPT has been developed into a practical and effective tool for calculation of noncovalent interactions and their interpretation.~\cite{Hohenstein:12,Szalewicz:12,Jansen:14} In spite of the growing popularity of the method, the existing many-electron formulations are limited to the single-reference description of the monomers. This work addresses the problem of extending SAPT to interactions between molecules which feature significant static correlation effects and cannot be described by a single Slater determinant.
We introduce and discuss a new multireference approach to obtain the second-order exchange energy contribution responsible for damping of the dispersion energy.

In the second order of SAPT it is natural to distinguish exchange-induction and exchange-dispersion contributions. 
At the van der Waals minimum the exchange-induction energy dampens a substantial part of its induction counterpart, while the exchange-dispersion component quenches only up to 15\% of the dispersion energy. In spite of that, exchange-dispersion component cannot be neglected if a quantitative description of a dimer is needed. 
Semiclassical and nonlocal models of the dispersion interaction represent the exchange-dispersion contribution using a model damping function, e.g., of the Tang-Toennis form.~\cite{Grimme:16,Tkatchenko:17,Shahbaz:19}
In many-body aggregates the nonadditive, three-body part of exchange-dispersion energy plays an important role, as it may be comparable to or even larger than the three-body dispersion contribution.~\cite{Podeszwa:07,Deible:14,Hapka:17} An illustrative example is the argon crystal where nonadditive exchange-dispersion effects differentiate between the face-centered cubic and hexagonal closed-packed structures.~\cite{Lotrich:97}

First direct calculations of the exchange-dispersion energy were performed for the \ce{H2+} ion\cite{Chalbie:73a,Chipman:73} and for the helium dimer.~\cite{Chalbie:76} In 1977 Cha{\l}asi{\'n}ski and Jeziorski\cite{Chalbie:77b} derived the explicit many-electron formula for the interaction of closed-shell monomers in terms of pair functions (the orbital version was given later by Rybak~\textit{et al}.\cite{Rybak:91}). Since the pioneering work on the Be-Be dimer,~\cite{Chalbie:77b} many-body SAPT (MB-SAPT)\cite{Szalewicz:12} calculations of the exchange-dispersion correction routinely invoke several important approximations.
First, electron correlation effects within the noninteracting monomers are neglected which is equivalent to the Hartree-Fock-based SAPT (HF-SAPT) approach. Second, calculations are usually performed at the uncoupled level of theory, i.e., without accounting for the change of monomer orbitals due to the perturbation field of the interacting partner. Finally, only terms up to the second order in the intermonomer overlap are kept in the exchange energy expression, which is referred to as the single-exchange or the $S^2$ approximation.~\cite{Murrell:65}

Inclusion of intramonomer correlation effects in the exchange-dispersion contribution is currently possible in two different SAPT flavors. DFT-SAPT\cite{Hesselmann:05} and the equivalent SAPT(DFT)\cite{Misquitta:05} approaches combine Kohn-Sham description of the interacting monomers with response properties from time-dependent Kohn-Sham (TD-KS) equations. Although one-electron reduced density matrices required in SAPT exchange energy calculations cannot be accurately reproduced at the Kohn-Sham level of theory even in the exact exchange-correlation functional limit,~\cite{Jansen:01} numerical experience has proven that their use provides exchange energy components which closely match available benchmarks for single-reference systems.~\cite{Korona:09,Korona:13} Open-shell exchange-dispersion formulas valid for both Kohn-Sham and Hartree-Fock treatment of high-spin systems also exist in the literature.~\cite{PZuch:08,Hapka:12,Gonthier:16}
An alternative way to account for intramonomer correlation is the SAPT(CC) method \cite{Korona:08a,Korona:08b,Korona:09} developed by Korona. SAPT(CC) is based on coupled cluster monomer wave functions limited to single and double excitations (CCSD) and response properties obtained from time-independent CCSD\cite{Moszynski:05} calculations. In her work on the exchange-dispersion component,~\cite{Korona:09} Korona introduced a new expression in terms of monomer density-matrix susceptibilities and density matrices.
This allowed her not only to derive the SAPT(CC) variant, but also to give theoretical background for the coupled approach. The latter had been proposed earlier by Hesselmann and co-authors\cite{Hesselmann:05} in the DFT-SAPT context, yet rigorous derivation had been missing. In general, the quality of exchange-dispersion energies based on transition density matrices from coupled Kohn-Sham or coupled Hartree-Fock calculations is significantly better when compared to the uncoupled result.~\cite{Korona:09}

Second-order exchange energies correct through all orders in intermolecular overlap may be obtained for single-determinant ground-state wave functions following the approach by Sch{\"a}ffer and Jansen.~\cite{Schaffer:12,Schaffer:13} 
Their DFT-SAPT results showed that the $S^2$ approximation tends to slightly overestimate the exchange-dispersion energy for intermediate distances and breaks down at short intermonomer separation, where this contribution may become negative.~\cite{Schaffer:13} 

One of the ongoing challenges for SAPT is the extension towards many-electron molecules which demand a multireference treatment. Recently, we have described a method for calculating second-order dispersion energy in such systems.~\cite{Hapka:19} The proposed approach is based on response properties from extended random phase approximation (ERPA)\cite{Pernal:12,Pernal:14b} equations. In this work we adopt the ERPA framework to derive an expression for the second-order SAPT exchange-dispersion energy. We present a spin-summed natural-orbital formula in the $S^2$ approximation applicable to both ground- and excited states of the monomers in singlet spin states. Our ERPA-based expression scales with the sixth-power of the size of the system and may be applied with any multireference method which provides one- and two-electron reduced density matrices. We consider wave functions of two kinds: the complete active space (CAS) reference and generalized valence bond perfect pairing (GVB) ansatz. Both coupled and uncoupled approximations are compared and discussed.

Another important step in the direction of a multireference SAPT formulation has recently been made by Patkowski and co-workers\cite{Patkowski:18,Waldrop:19} who devised a spin-flip variant of SAPT (SF-SAPT) to calculate first-order exchange energy correction for an arbitrary spin state of the dimer. In their method a multireference low-spin state of the complex is approached starting from high-spin restricted open-shell Hartree-Fock (ROHF) monomer wave functions. The original SF-SAPT paper\cite{Patkowski:18} presented a first-order exchange formula limited to single-exchange of electrons. Subsequently, Waldrop and Patkowski\cite{Waldrop:19} were able to move beyond the $S^2$ limit and introduced a milder and more accurate single-spin-flip approximation based on the ideas of Sch{\"a}fer and Jansen.

The outline of the paper is as follows: In Section~\ref{sec:th} the exchange-dispersion energy is expressed in terms of one- and two-electron reduced density matrices from ERPA calculations. Both the scaling behavior of the introduced multireference variant as well as its relation to the single reference case are briefly discussed. Section~\ref{sec:compdet} contains the relevant details regarding our implementation and computations. Numerical demonstration for representative multi- and single-reference-dominated dimers is given in Section~\ref{sec:res}. Section~\ref{sec:concl} concludes our study.

\section{Theory \label{sec:th}}
We consider interaction of two monomers A and B in states $\ket{\Psi^A_I}$ and $\ket{\Psi^B_J}$, respectively, with the corresponding energies $E^A_I$ and $E^B_J$. Consequently the zeroth-order wave function of a dimer reads $\ket{\Psi^{(0)}}=\ket{\Psi^A_I\Psi^B_J}$. Indices $I$ and $J$ do not necessarily correspond to ground states, i.e., one or both of the monomers may be in electronically excited state. The only assumption in the presented formalism is that $\ket{\Psi^{(0)}}$ is nondegenerate. 

The general expression for the second-order exchange-dispersion term in the symmetrized Rayleigh-Schr{\"o}dinger theory (SRS) takes the following form:
\begin{equation}
\begin{split}
E^{(2)}_{\rm exch-disp} &= 
\frac{\langle \Psi_I^A\Psi_J^B | \left( \hat{V} - E^{(1)}_{\rm SRS} \right) \hat{\mathcal{A}} \Psi^{(1)}_{\rm disp} \rangle}{\braket{\Psi_I^A\Psi_J^B|\hat{\mathcal{A}}\Psi_I^A\Psi_J^B}} \\
&- \braket{\Psi_I^A\Psi_J^B|\hat{V}\Psi^{(1)}_{\rm disp}} \ \ \ ,
\end{split}
\label{eq:exdF}
\end{equation}
where the intermolecular interaction operator $\hat{V}$ represents all Coulomb interactions between electrons and nuclei of the two monomers, $\hat{\mathcal{A}}$ is the antisymmetrizer that exchanges electrons between the monomers, $E^{(1)}_{\rm SRS}$ denotes the first-order energy in SRS, and $\ket{\Psi^{(1)}_{\rm disp}}$ is the first-order dispersion wave function
\begin{equation}
\Psi^{(1)}_{\rm disp} = \sum_{\mu \neq I, \nu\neq J } \frac{\ket{\Psi^A_\mu\Psi^B_\nu}\braket{\Psi^A_\mu\Psi^B_\nu|\hat{V}|\Psi_I^A\Psi_J^B}}{E_I^A + E_J^B - E_\mu^A - E_\nu^B} \ \ \ , 
\label{eq:disp1}
\end{equation}
so the last term in Eq.~\eqref{eq:exdF} denotes the second-order dispersion energy $E^{(2)}_{\rm disp}=\braket{\Psi_I^A\Psi_J^B|\hat{V}\Psi^{(1)}_{\rm disp}}$.
Greek indices used in Eq.~\eqref{eq:disp1} and throughout the paper pertain to electronic states of monomers.

In the $S^2$ approximation\cite{Murrell:65} one restricts the antisymmetrizer in Eq.~\eqref{eq:exdF} to single-exchange of electrons between the monomers: $\hat{\mathcal{A}} \approx \hat{1} + \hat{P}_1$, with $\hat{P}_1$ being the sum of all permutations, $\hat{P}_{ij}$, interchanging the coordinates of electrons $i$ and $j$
\begin{equation}
\hat{P}_1 = -\sum_{i=1}^{N_A}\sum_{j=1+N_A}^{N_A+N_B} \hat{P}_{ij} \ \ \ , 
\end{equation}
$N_A$ and $N_B$ denoting the number of electrons in pertinent monomers. The exchange-dispersion energy in this approximation reads:\cite{Chalbie:77a,Chalbie:77b}
\begin{equation}
\begin{split}
E^{(2)}_{\rm exch-disp} &= \langle \Psi_I^A\Psi_J^B | \hat{V}\hat{P}_1 \Psi^{(1)}_{\rm disp} \rangle \\
&- E^{(1)}_{\rm elst}\langle  \Psi_I^A\Psi_J^B | \hat{P}_1 \Psi^{(1)}_{\rm disp} \rangle \\
&- E^{(2)}_{\rm disp}\langle  \Psi_I^A\Psi_J^B | \hat{P}_1 \Psi_I^A\Psi_J^B \rangle \ \ \ , 
\label{eq:exds2}
\end{split}
\end{equation}
where $E^{(1)}_{\rm elst}=\braket{\Psi_I^A\Psi_J^B|\hat{V}|\Psi_I^A\Psi_J^B}$ is the first-order electrostatic energy.

Evaluation of Eq.~\eqref{eq:exds2} requires access to one- and two-electron transition reduced density matrices of the monomers (1-TRDMs and 2-TRDMs, respectively). In this work we focus on multireference description of the system and explore the possibility to obtain the necessary response properties from Rowe's equation of motion (REOM) theory\cite{Rowe:68} in the extended random phase approximation.~\cite{Pernal:12} 
The 1-TRDM connecting a state of interest $\ket{\Psi}$ with a higher- or lower state $\ket{\Psi_\nu}$, 
\begin{equation}
\gamma_{pq}^{\nu}=\braket{\Psi|\hat{a}_q^{\dagger}\hat{a}_p|\Psi_{\nu}} \ \ \ ,
\label{eq:def1trdm}
\end{equation}
has the following form in the REOM-ERPA framework:~\cite{Pernal:12,Pernal:14}
\begin{align}
\forall_{p>q}\ \ \ \gamma_{pq}^{\nu} &  =-(n_{p}-n_{q})\left[  \mathbf{X}_{\nu
}\right]_{pq} \ \ \ , \\
\forall_{p<q}\ \ \ \gamma_{pq}^{\nu} &  =-(n_{p}-n_{q})\left[  \mathbf{Y}_{\nu
}\right]_{qp}\ \ \ ,
\label{eq:1trdm}
\end{align}
where we assumed natural spinorbital representation of the wave function $\ket{\Psi}$. Throughout the text  the indices $pqrs$ pertain to natural orbitals of  monomers. The formalism is developed for monomers in singlet states so the natural occupation numbers for $\alpha$ and $\beta$ spin-blocks of one-electron reduced density matrix (1-RDM) are equal and $\forall_p \ \ 0 \le n_p \le 1$, the occupation numbers sum up to  half a number of electrons, $\sum_p n_p = N/2$. 
Vectors $\left[ \mathbf{X}_{\nu}, \mathbf{Y}_{\nu} \right]$ correspond to solutions of the the generalized ERPA eigenproblem:~\cite{Pernal:14}
\begin{equation}
\left(
\begin{array}
[c]{cc}%
\mathbf{\mathcal{A}} & \mathbf{\mathcal{B}}\\
\mathbf{\mathcal{B}} & \mathbf{\mathcal{A}}%
\end{array}
\right)  \left(
\begin{array}
[c]{c}%
\mathbf{X}_{\nu}\\
\mathbf{Y}_{\nu}%
\end{array}
\right)  =\omega_{\nu}\left(
\begin{array}
[c]{cc}%
-\mathbf{\mathcal{N}} & \mathbf{0}\\
\mathbf{0} & \mathbf{\mathcal{N}}%
\end{array}
\right)  \left(
\begin{array}
[c]{c}%
\mathbf{X}_{\nu}\\
\mathbf{Y}_{\nu}%
\end{array}
\right)  \ \ \ ,\label{ERPA}%
\end{equation}
with the metric matrix expressed through the natural occupation numbers $\left\{  n_{p}\right\}$:
\begin{equation}
\forall_{\substack{p>q\\r>s}} \ \ \ \mathcal{N}_{pq,rs}=(n_{p}-n_{q})\delta_{pr}\delta_{qs}\ \ \ .
\label{eq7}
\end{equation}
Eigenvalues $\{\omega_\nu\}$ provide approximations to electronic excitation energies of a system.
The ERPA equations are solved independently for each of the monomers (in Eq.~\eqref{eq:def1trdm}-\eqref{eq7} the monomer index was dropped for convenience). For a given monomer Hamiltonian $\hat{H}$ and a reference wave function $\ket{\Psi}$
the matrices $\mathbf{\mathcal{A}}$ and $\mathbf{\mathcal{B}}$ are defined as 
$\left[ \mathbf{\mathcal{A}} \right]_{pqrs}=\left[\mathbf{\mathcal{B}} \right]_{pqsr}=
\left\langle \Psi |[\hat{a}_{p}^{\dag}\hat{a}_{q},[\hat
{H},\hat{a}_{s}^{\dag}\hat{a}_{r}]] | \Psi \right\rangle$ and are expressed exclusively in terms of one- and two-electron reduced density matrices of the system (for their explicit GVB- and CAS-based forms see Refs.~\onlinecite{Pernal:14b} and \onlinecite{Pastorczak:18a}, respectively). The reference wavefunction of a monomer pertains to a state of interest $I$. In a special case when $\ket{\Psi}$ is a single determinant the ERPA equations turn into those of TD-HF. 

If the monomer is in the ground state, it is convenient to formulate ERPA equations as a half-sized real-value symmetric eigenproblem making use of the fact the $\mathcal{A}+\mathcal{B}$ and $\mathcal{A}-\mathcal{B}$ matrices (the electronic Hessian matrices) are positive definite.
When the monomer wave function represents a system in the excited state, we solve a nonsymmetric real-value eigenproblem, as described in Ref.~\onlinecite{Pastorczak:18}.

Application of the ERPA approach together with a CAS or GVB multireference wave function involves partitioning the orbital space into three disjoint subsets denoted $s_1$, $s_2$ and $s_3$ \cite{Hapka:19}. In the case of a CAS reference $s_1$, $s_2$ and $s_3$ correspond to the inactive, active and virtual orbitals, respectively. For the GVB wave function, the $s_1$ set contains orbitals with occupation numbers $n_p > 0.992$. The $s_2$ set includes orbitals which satisfy the $0.992 \geq n_p \geq 0.5$ condition together with their ``weakly'' occupied partners which are coupled to form geminals. The $s_3$ set hosts the remaining orbitals with occupation numbers $n_p < 0.5$. The $p, q$ indices of the ERPA matrices belong to the following subspaces:
\begin{equation}
\begin{matrix}
p \in s_2 \wedge q \in s_1 \\
p \in s_3 \wedge q \in s_1 \\
p \in s_2 \wedge q \in s_2 \\
p \in s_3 \wedge q \in s_2, \\
\end{matrix}
\end{equation}
with $r, s$ indices spanning the equivalent range. In the following we will use $M_{s_1}$, $M_{s_2}$ and $M_{s_3}$ symbols to represent the cardinalities of the orbital subsets.

In all derivations we assume spin-preserving adaptation of the ERPA excitation operator, which is equivalent to imposing the $[\mathbf{X}_{\nu}]_{p_{\alpha}q_{\alpha}}$ = $[\mathbf{X}_{\nu}]_{p_{\beta}q_{\beta}}$ and $[\mathbf{X}_{\nu}]_{p_{\alpha}q_{\beta}}$ = $[\mathbf{X}_{\nu}]_{p_{\beta}q_{\alpha}}$ = 0 restrictions on eigenvectors (same holds for $[\mathbf{Y_{\nu}}]$).
Consequently, only the $\alpha\alpha$ and $\beta\beta$ blocks of the 1-TRDM are nonzero. Additionally, for singlet states $\alpha\alpha$ and $\beta\beta$ blocks are equal: $\gamma^{\nu}_{p_{\alpha}q_{\alpha}}$ = $\gamma^{\nu}_{p_{\beta}q_{\beta}}$ = $\gamma^{\nu}_{pq}$. 

An important step in  deriving the expression for the exchange-dispersion energy is finding the ERPA form of the 2-TRDM defined as
\begin{equation}
\Gamma^{\nu}_{pqrs}=\braket{\Psi|\hat{a}_r^{\dagger}\hat{a}_s^{\dagger}\hat{a}_q\hat{a}_p|\Psi_{\nu}} \ \ \ . 
\end{equation}
This can be achieved within the REOM theory, in which transition matrix elements of a given operator $\hat{F}$ are accessed as expectation values of a commutator of $\hat{F}$ and the excitation operator $\hat{O}^{\dagger}_\nu$:
\begin{equation}
\langle \Psi | \hat{F} | \Psi_{\nu} \rangle = \langle \Psi | \big[ \hat{F}, \hat{O}^{\dagger}_{\nu} \big] | \Psi \rangle  \ \ \ ,
\label{eq:comm}
\end{equation} 
and it is assumed that the excitation operator creates $\ket{\Psi_\nu}$ as $\hat{O}^{\dagger}_\nu \ket{\Psi} = \ket{\Psi_\nu}$, while the deexcitation operator $\hat{O}_{\nu}$ satisfies the condition $\hat{O}_\nu\ket{\Psi} = 0$. The combination of ERPA excitation operator truncated to single excitations
\begin{equation}
\hat{O}_{\nu}^{\dag}=\sum_{p>q}\left[  \mathbf{X}_{\nu}\right]  _{pq}\hat
{a}_{p}^{\dag}\hat{a}_{q}+\sum_{p>q}\left[  \mathbf{Y}_{\nu}\right]  _{pq}%
\hat{a}_{q}^{\dag}\hat{a}_{p} \ \ \ ,
\end{equation}
with $\hat{F}=\hat{a}_r^{\dagger}\hat{a}_s^{\dagger}\hat{a}_q\hat{a}_p$ leads to the half of spin-summed expression for 2-TRDM:
\begin{equation}
\begin{split}
\bar{\Gamma}^{\nu}_{pqrs} &= \Gamma^{\nu}_{p_{\alpha}q_{\alpha}r_{\alpha}s_{\alpha}} +   
                             \Gamma^{\nu}_{p_{\beta}q_{\alpha}r_{\beta}s_{\alpha}} \\
& =\sum_{t<p}[\mathbf{X}_\nu]_{pt}\bar{\Gamma}_{tqrs} + \sum_{t<q}[\mathbf{X}_\nu]_{qt}\bar{\Gamma}_{ptrs} \\
          & - \sum_{t>r}[\mathbf{X}_\nu]_{tr}\bar{\Gamma}_{pqts} - \sum_{t>s} [\mathbf{X}_\nu]_{ts}\bar{\Gamma}_{pqrt} \\
          & + \sum_{t>p}[\mathbf{Y}_\nu]_{tp}\bar{\Gamma}_{tqrs} + \sum_{t>q} [\mathbf{Y}_\nu]_{tq}\bar{\Gamma}_{ptrs} \\
          & - \sum_{t<r}[\mathbf{Y}_\nu]_{rt}\bar{\Gamma}_{pqts} - \sum_{t<s} [\mathbf{Y}_\nu]_{st}\bar{\Gamma}_{pqrt} \ \ \ , \\
\end{split}
\label{eq:2trdm}
\end{equation}
where $\bar{\Gamma}_{pqrs}$ is the half of the spin-summed 2-RDM, $\Gamma_{pqrs}=\braket{\Psi|\hat{a}^{\dagger}_r\hat{a}^{\dagger}_s \hat{a}_q\hat{a}_p|\Psi}$,
i.e., $\bar{\Gamma}_{pqrs} = \Gamma_{p_{\alpha}q_{\alpha}r_{\alpha}s_{\alpha}} + \Gamma_{p_{\beta}q_{\alpha}r_{\beta}s_{\alpha}}$. Analogous expression holds for the $\beta\beta\beta\beta + \alpha\beta\alpha\beta$ component of 2-TRDM. 
In subsequent derivations we make use of the fact that both monomers are in spin-singlet states (pertinent $\alpha$ and $\beta$ blocks of TRMDs are equal) and that  both the exact 2-TRDM and its ERPA approximation, Eq.(\ref{eq:2trdm}), satisfy the sum-rule
\begin{equation}
\sum_p \Gamma^{\nu}_{pqps} = (N-1)\gamma^{\nu}_{qs} \ \ \ ,
\end{equation}
where $N$ is a number of electrons. 

From now on, indices $ijkl$ pertain to atomic orbitals $\left\{ \chi(\mathbf{r}) \right\}$, whereas, as it has been already indicated, indices $pqrs\ldots$ refer to natural orbitals $\left\{ \varphi(\mathbf{r}) \right\}$. The effective two-electron potential $\tilde{v}(\mathbf{r},\mathbf{r}')$ has the form:
\begin{equation}
\tilde{v}(\mathbf{r},\mathbf{r}') = |\mathbf{r} - \mathbf{r}'|^{-1 } + \frac{1}{N_B}v^B(\mathbf{r}) + \frac{1}{N_A}v^A(\mathbf{r}') \ \ \ ,
\end{equation}
with $v^X(\textbf{r}) = -\sum_{\alpha \in X} Z_{\alpha}|\mathbf{r}-\mathbf{R}_{\alpha}|^{-1}$, where the summation runs over all nuclei of monomer $X$. Its matrix representation is
\begin{equation}
\begin{split}
\tilde{v}_{pq}^{rs} &= \bigg \langle \varphi_p(\mathbf{r})\varphi_q(\mathbf{r}') | \tilde{v}(\mathbf{r},\mathbf{r}') | \varphi_r(\mathbf{r})\varphi_s(\mathbf{r}') \bigg \rangle \\
&= v_{pq}^{rs} + N_B^{-1}\braket{\varphi_p|v^B|\varphi_r}\big(\delta_{qs}\delta_{X_q X_s} + S_q^s (1-\delta_{X_q X_s})\big) \\
&+ N_A^{-1}\braket{\varphi_q|v^A|\varphi_s}\big( \delta_{pr}\delta_{X_p X_r} + S_p^r(1-\delta_{X_p X_r}) \big) \ \ \ ,
\end{split}
\end{equation}
where a given orbital $\varphi_p$ may belong either to monomer $X_p = A$ or $X_p = B$, $S_p^q$ is the overlap integral $S_p^q=\braket{\varphi_p|\varphi_q}$. Notice that natural orbitals belonging to the same monomer are orthonormal and $\forall_{pq\in X} \ S_p^q=\delta_{pq}$. A regular two-electron Coulomb integral is denoted as $v_{pq}^{rs}=\langle \varphi_p(\mathbf{r}_1)\varphi_q(\mathbf{r}_2) | r_{12}^{-1} | \varphi_r(\mathbf{r}_1)\varphi_s(\mathbf{r}_2)  \rangle$.

The exchange-dispersion energy is size-extensive, provided that N-representable RDMs and size-extensive TRDMs are employed. Therefore, the last two disconnected terms in Eq.~\eqref{eq:exds2} should cancel with contributions from the first term. Korona introduced transition density cumulants\cite{Korona:08b,Korona:09} to explicitly carry out the cancellation. Since we do not exploit the cumulant expansion, the disconnected terms have to be kept.

When Eq.~\eqref{eq:exds2} is rewritten in terms of transition density matrices, it is straightforward to recognize its structure within the ERPA framework. This leads to the following formula:
\begin{equation}
\begin{split}
E^{(2)}_{\rm exch-disp} &= 8\sum_{\mu,\nu}\frac{D_{\mu\nu}s_{\mu\nu}}{\omega_\mu^A + \omega_\nu^B} \\
 &- 8\left(E^{(1)}_{\rm elst} - V^{AB} \right)\sum_{\mu,\nu} \frac{s_{\mu\nu}t_{\mu\nu}}{\omega_\mu^A + \omega_\nu^B} \\
 &+ 2 E^{(2)}_{\rm disp}\sum_{ijkl}\gamma^A_{ij}\gamma^B_{kl}S_i^l S_j^k \ \ \ .
\label{eq:exdmat}
\end{split}
\end{equation}
where $V^{AB}$ is the internuclear repulsion energy. The last term includes 1-RDMs of monomers in the AO basis, namely ${\bf{\gamma} }^X_{AO}=[{\bf C}^X]^T {\bf n}^X {\bf C}^X$, where ${\bf C}^X$ transform atomic orbitals to the natural orbitals of the monomer $X$ and ${\bf n}^X$ is a vector of the natural occupation numbers of $X$.
The electrostatic energy that appears in the second term in the AO basis reads
\begin{equation}
\begin{split}
E^{(1)}_{\rm elst} &= 2 \sum_{ij} \gamma^A_{ij}v^B_{ij} + 2 \sum_{ij} \gamma^B_{ij}v^A_{ij} \\
&+ 4 \sum_{ijkl} \gamma^A_{ij}\gamma^B_{kl} v_{ik}^{jl} + V^{AB} \ \ \ .
\end{split}
\end{equation}
The dispersion energy in ERPA takes the form:~\cite{Jaszunski:85,Hapka:19} 
\begin{equation}
\begin{split}
E^{(2)}_{\rm disp} &= -16 \sum_{\mu, \nu} \frac{\Big(\sum_{\substack{pq \in A \\ rs \in B}} \gamma^{A,\mu}_{pq}\gamma^{B,\nu}_{rs} v_{pr}^{qs} \Big)^2}{\omega^A_\mu + \omega^B_\nu} \\
&= -16 \sum_{\mu,\nu }\frac{s_{\mu\nu}^2}{\omega^A_\mu + \omega^B_\nu}  \ \ \ ,
\end{split}
\end{equation}
where $\omega^{A/B}_{\mu/\nu}$ denotes transition energies obtained as eigenvalues of the ERPA problem, Eq.~(\ref{ERPA}), solved separately for each monomer,  1-TRDM is obtained as shown in Eq.~(\ref{eq:1trdm}), and the matrix ${\bf s}$ employed also in Eq.~\eqref{eq:exdmat} is defined as 
\begin{equation}
\begin{split}
s_{\mu\nu} &= \sum_{\substack{p>q \in A \\ r>s \in B}}\Big([\textbf{Y}_\mu^A]_{pq}-[\textbf{X}_\mu^A]_{pq})([\textbf{Y}_\nu^B]_{rs}-[\textbf{X}_\nu^B]_{rs}\Big) \\ 
&\times(n_p-n_q)(n_r-n_s) v_{pr}^{qs} \ \ \ .
\end{split}
\end{equation}

The remaining intermediates in Eq.~\eqref{eq:exdmat} read 
\begin{equation}
\begin{split}
t_{\mu\nu} &= \sum_{\substack{pq \in A \\ rs \in B}} \gamma^{A,\mu}_{pq}\gamma^{B,\nu}_{rs} S_p^s S_q^r \\
&= \sum_{\substack{p>q \in A \\ r>s \in B}}\Big( [\mathbf{X}_\mu^A]_{pq}[\mathbf{X}_\nu^B]_{rs} + [\mathbf{Y}_\mu^A]_{pq}[\mathbf{Y}_\nu^B]_{rs} \Big) \\ &\times (n_p-n_q)(n_r-n_s) S_p^s S_q^r \\
&- \sum_{\substack{p>q \in A \\ r>s \in B}} \Big( [\mathbf{Y}_\mu^A]_{pq}[\mathbf{X}_\nu^B]_{rs} + [\mathbf{X}_\mu^A]_{pq}[\mathbf{Y}_\nu^B]_{rs} \Big) \\ &\times (n_p-n_q)(n_r-n_s) S_p^r S_q^s \\
\end{split}
\end{equation}
and
\begin{equation}
\begin{split}
D_{\mu\nu} &= \sum_{\substack{pq \in A \\ rs \in B }} \gamma^{A,\mu}_{pq}\gamma^{B,\nu}_{rs} \tilde{v}_{qp}^{rs} 
+ \sum_{\substack{pq \in A \\ rsbb' \in B}} \gamma^{A,\mu}_{pq}\bar{\Gamma}^{B,\nu}_{rsbb'} S_p^{b'} \tilde{v}_{qb}^{sr} \\
&+ \sum_{\substack{ pqaa' \in A \\ rs \in B }} \gamma^{B,\nu}_{rs}\bar{\Gamma}^{A,\mu}_{pqaa'} S_{a'}^r \tilde{v}_{pq}^{as} \\
&+\sum_{\substack{pqaa' \in A \\ rsbb' \in B}} \bar{\Gamma}^{A,\mu}_{pqaa'}\bar{\Gamma}^{B,\nu}_{rsbb'} S_q^{b'} S_{a'}^s \tilde{v}_{pr}^{ab}  \ \ \ . \\
\end{split}
\end{equation}

Expansion of 1- and 2-TRDMs according to Eq.~\eqref{eq:1trdm} and Eq.~\eqref{eq:2trdm}, respectively, unfolds the structure of the $D_{\mu\nu}$ matrix:
\begin{equation}
\begin{split}
  D_{\mu\nu} &= \sum_{\substack{p>q \in A \\ r>s \in B}} [\mathbf{X}^A_\mu]_{pq}[\mathbf{X}^B_\nu]_{rs}
  \Big[
    (n_p-n_q)(n_r-n_s) \tilde{v}_{qp}^{rs} \\
  &  +(n_r-n_s) P^A_{pqrs}
    +(n_p-n_q) P^B_{pqrs}
    -P^{AB}_{pqrs}
    \Big] \\
    & + \sum_{\substack{p>q \in A \\ r>s \in B}} [\mathbf{Y}^A_\mu]_{pq}[\mathbf{Y}^B_\nu]_{rs}
  \Big[
  (n_p-n_q)(n_r-n_s) \tilde{v}_{pq}^{sr} \\
  & -(n_r-n_s) P^A_{qpsr}
  -(n_p-n_q) P^B_{qpsr}
  -P^{AB}_{pqrs}  
  \Big] \\
    & + \sum_{\substack{p>q \in A \\ r>s \in B}} [\mathbf{X}^A_\mu]_{pq}[\mathbf{Y}^B_\nu]_{rs}
  \Big[
   -(n_p-n_q)(n_r-n_s) \tilde{v}_{qp}^{sr} \\
  & -(n_r-n_s) P^A_{pqsr}
  +(n_p-n_q) P^B_{pqsr}
  - P^{AB}_{pqsr}
  \Big] \\ 
    & + \sum_{\substack{p>q \in A \\ r>s \in B}} [\mathbf{Y}^A_\mu]_{pq}[\mathbf{X}^B_\nu]_{rs}
  \Big[
  -(n_p-n_q)(n_r-n_s) \tilde{v}_{pq}^{rs} \\
  & +(n_r-n_s) P^A_{qprs}
  -(n_p-n_q) P^B_{qprs}
  - P^{AB}_{pqsr}
  \Big] \\ 
  \label{eq:Dmat}
\end{split}
\end{equation}
where we have introduced $P^X$ matrices defined as 
\begin{equation}
\begin{split}
 P^A_{pqrs} &= \sum_{aa' \in A} \big( 
 N^A_{aa'pr} \tilde{v}_{qa'}^{as}
 - N^A_{qa'ar} \tilde{v}_{pa'}^{as}
 - N^A_{aqa'r} \tilde{v}_{a'p}^{as}\big) \\
& + O^A_{ps} S_q^r \\
\end{split}
\end{equation}
\begin{equation}
\begin{split}
P^B_{pqrs} &= \sum_{bb' \in B} \big(
  N^B_{bb'rp} \tilde{v}_{qb}^{b's}
- N^B_{sb'bp} \tilde{v}_{qb}^{b'r} 
- N^B_{bsb'p} \tilde{v}_{qb}^{rb'}\big) \\
&+ O^B_{rq} S_p^s \\
\end{split}
\end{equation}
\begin{equation}
  \begin{split}
  P^{AB}_{pqrs} &= \sum_{\substack{a \in A\\b \in B}} \Big( 
   T_{aqbs} \tilde{v}_{pr}^{ab} + T_{aqrb} \tilde{v}_{pb}^{as} \\
 & + T_{pabs} \tilde{v}_{ar}^{qb} + T_{parb} \tilde{v}_{ab}^{qs} \\
 & + U^A_{aqbs} V^B_{parb} + U^A_{aqrb} V^B_{pabs} \\
 & + U^A_{pabs} V^B_{aqrb}   + U^A_{parb} V^B_{aqbs} \\
 & + V^A_{aqbs} U^B_{parb} + V^A_{aqrb} U^B_{pabs} \\
 & + V^A_{pabs} U^B_{aqrb}   + ^A_{parb} U^B_{aqbs} \\
 & + W_{aqbs}S_p^bS_a^r + W_{aqrb}S_p^sS_a^b \\
 & + W_{pabs}S_a^bS_q^r + W_{parb}S_a^sS_q^b \Big)
\end{split}
\label{eq:pab}
\end{equation}
with the following intermediates
\begin{equation}
  N^A_{tuvw} = \sum_{a \in A} \bar{\Gamma}^A_{tuva} S_a^w, \quad
  N^B_{tuvw} = \sum_{b \in B} \bar{\Gamma}^B_{tuvb} S_w^b
\end{equation}
\begin{equation}
  O^A_{tu} = \sum_{aa'a'' \in A} \bar{\Gamma}^A_{taa'a''} \tilde{v}_{aa'}^{a''u}, \quad
  O^B_{tu} = \sum_{bb'b'' \in B} \bar{\Gamma}^B_{tbb'b''} \tilde{v}_{ub''}^{b'b}
\end{equation}
\begin{equation}
  U^A_{tuvw} = \sum_{aa' \in A} \bar{\Gamma}^A_{taua'}S_{a}^v S_{a'}^w, \,
  U^B_{tuvw} = \sum_{bb' \in B} \bar{\Gamma}^B_{bvb'w} S_t^{b} S_u^{b'}
\end{equation}
\begin{equation}
  V^A_{tuvw} = \sum_{aa' \in A} \bar{\Gamma}^A_{taua'}\tilde{v}_{av}^{a'w}, \quad
  V^B_{tuvw} = \sum_{bb' \in B} \bar{\Gamma}^B_{bvb'w} \tilde{v}_{tb}^{ub'}
\end{equation}
\begin{equation}
\begin{split}
  T_{tuvw} &= \sum_{\substack{aa' \in A \\ bb' \in B }} \bar{\Gamma}^A_{taua'}\bar{\Gamma}^B_{bvb'w} S_{a'}^{b} S_{a}^{b'} \\
  &= \sum_{bb' \in B} \bigg(\sum_{aa' \in A} \bar{\Gamma}^A_{taua'} S_{a}^{b'} S_{a'}^{b}\bigg)\bar{\Gamma}^B_{bvb'w} \\
  &=\sum_{bb' \in B }U^A_{tub'b}\bar{\Gamma}^B_{bvb'w}
\end{split}
\end{equation}
\begin{equation}
\begin{split}
  W_{tuvw} &= \sum_{\substack{aa' \in A \\ bb' \in B }} \bar{\Gamma}^A_{taua'}\bar{\Gamma}^B_{bvb'w} \tilde{v}_{ab}^{a'b'} \\
  &= \sum_{bb' \in B} \bigg(\sum_{aa' \in A} \bar{\Gamma}^A_{taua'}\tilde{v}_{ab}^{a'b'}\bigg)\bar{\Gamma}^B_{bvb'w} \\
  &=\sum_{bb' \in B}V^A_{tubb'}\bar{\Gamma}^B_{bvb'w}
  \label{eq:intW}
\end{split}
\end{equation}
(indices in all intermediates correspond to natural orbitals of the pertinent monomers).

The expression for the exchange-dispersion energy, Eq.(\ref{eq:exdmat}), has been proposed primarily to use it together with the multireference description of monomers. Naturally, it is also applicable when both monomers are described with single determinants.
As it has already been mentioned, in such a case the ERPA equations become equivalent to the TD-HF approach, so the resulting density-matrix-based expression\cite{Moszynski:94} for the exchange-dispersion energy could be called ``coupled'' HF. It should be noticed that it would not be identical to the ``coupled'' HF equation proposed by Hesselmann \textit{et al}.\ in Eq.(20) in Ref.~\onlinecite{Hesselmann:05}. The latter was obtained by replacing the uncoupled amplitudes in the expression derived by applying second quantization formalism\cite{Moszynski:94b} with the coupled ones. Later, this approach gained theoretical justification from Korona,~\cite{Korona:09} who derived a generally valid formula written in terms of density matrices and frequency-dependent density- and density-matrix susceptibilities, and recovered results of Hesselmann \textit{et al}.\ by employing the same coupled HF amplitudes. Our expression could in principle also be turned into the coupled approach proposed by Hesselmann \textit{et al}.\ (implemented in \textsc{Molpro}\cite{Molpro:12}) by setting the $\mathbf{X}$ components of the ERPA eigenvectors to zero and replacing the $\mathbf{Y}$ components in Eq.~\eqref{eq:exdmat}-\eqref{eq:intW} by the pertinent $\mathbf{Y}-\mathbf{X}$ combinations. This is equivalent to the use of symmetrized 1-TRDMs from TD-HF or TD-KS equations (see Supplementary Information for details). We have exploited such replacements to test our code against the \textsc{Molpro} implementation for single reference cases. 
We have also verified that for the Hartree-Fock reference our approach and the one presented in Ref.~\onlinecite{Hesselmann:05} lead to nearly identical numerical results (see Table~S1 in the Supporting Information), if applied to systems which do not require multireference description. However, if used for multireference systems, the two formulations may give substantially different results, as we observed for the beryllium dimer (see Section~\ref{sec:res:model}). In such a situation there is no reason to trust more one result over the other.

Evaluation of Eq.~\eqref{eq:exdmat} scales with the sixth power of the molecular size. The construction of matrices $\mathbf{P}^A$, $\mathbf{P}^B$ and $\mathbf{P}^{AB}$ [Eq.~\eqref{eq:pa}-\eqref{eq:pab}] engages two four-index quantities (2-RDMs, integrals or intermediates) and has the $n_{\rm OCC}^4 n_{\rm SEC}^2$ cost, where $n_{\rm OCC} = M_{s_1} + M_{s_2}$ and $n_{\rm SEC} = M_{s_2} + M_{s_3}$. The evaluation of matrix $\mathbf{D}$ in Eq.~\eqref{eq:Dmat} scales as $n_{\rm OCC}^3 n_{\rm SEC}^3$.
Note that the diagonalization of the full ERPA eigenproblem also scales as  $n_{\rm OCC}^3 n_{\rm SEC}^3$.
By comparison, the single-reference density-matrix-based expression scales with the fifth power of the system size and involves steps with a $M_{s_1}^3 M_{s_3}^2$ scaling (the $s_2$ space is empty). 

It is possible to devise low-scaling approximations to the exchange-dispersion energy formula by employing the Dyall partitioning of the monomer Hamiltonian\cite{Dyall:95,Rosta:02} in ERPA equations. The response properties may be subsequently expanded in the coupling parameter $\alpha$ which connects the zeroth-order, noninteracting Hamiltonian and the fully interacting one.~\cite{Hapka:19} Truncation of this expansion at the zeroth order leads the so-called uncoupled approximation, denoted $E^{\rm UC}_{\rm exch-disp}$. The uncoupled formula is identical to the coupled one with ERPA eigenvectors and eigenvalues replaced by their approximations of the zeroth-order in $\alpha$ [this is analogous to the uncoupled formula for dispersion, see Eq.~(26) in Ref.~\onlinecite{Hapka:19}]. The improved scaling behavior in the uncoupled model comes from reduction in the dimensionality of the ERPA eigenproblem. In particular, the electronic hessian matrices take a block-diagonal form in the $\alpha=0$ limit, provided that the set of active orbitals is not empty.~\cite{Pastorczak:18a} Thus, one avoids diagonalization of the full hessian matrix which scales with the sixth power of the molecular size. Instead, if CAS wavefunction is employed one solves only a number of low-dimensional eigenproblems, the largest of which is the active-active block with a $M_{s_2}^6$ scaling. This is still beneficial, since in standard CAS calculations the number of active orbitals is considerably smaller than the size of the virtual space. In GVB-based calculations the zeroth-order hessian matrix takes diagonal form which reduces the scaling to $M_{s_2}^5$.

Similarly to what has been proposed for the dispersion energy in Ref.~\onlinecite{Hapka:19} a semi-coupled, i.e.,\ first-order in $\alpha$, approximation for the exchange-dispersion energy could be derived in a straightforward manner. The scaling of the latter would place in between of the fully-coupled and uncoupled approaches. In this paper we present only results following from the fully-coupled and uncoupled expressions.

\section{Computational details \label{sec:compdet}}

Second-order exchange-dispersion energies based on the HF, CASSCF, GVB and FCI treatment of the monomers were obtained with the in-house code. 
We used a developer version of the \textsc{Molpro} program\cite{Molpro:12} as a source of all necessary integrals as well as 1- and 2-RDMs for Hartree-Fock and CASSCF wave functions. The GVB calculations were performed in the Dalton program\cite{Dalton:13} and interfaced with our code.
Both GVB and CASSCF calculations presented in Section~\ref{sec:sref} used MP2 natural orbitals as the starting guess. For details on the GVB implementation see Ref.~\onlinecite{Pastorczak:19a}.

Second-order dispersion and exchange-dispersion energies based on time-independent CCSD\cite{Korona:06} response functions, denoted SAPT(CCSD),~\cite{Korona:08,Korona:09} were calculated in the \textsc{Molpro}\cite{Molpro:12} package. Note that all exchange components in SAPT(CCSD) are formulated within the $S^2$ approximation.

All calculations were performed using augmented correlation-consistent orbital basis sets of double- and triple-zeta quality (aug-cc-pV$X$Z, $X=$ D,T).~\cite{Dunning:89,Kendall:92} A larger, doubly-augmented d-aug-cc-pVTZ basis\cite{Woon:94} was chosen for the \ce{He}$\cdots$\ce{H2} complex. Monomer calculations were performed in the dimer-centered basis set. 

The benchmark second-order exchange-dispersion energies at the FCI level of theory for the \ce{H2}$\cdots$\ce{H2} and \ce{He}$\cdots$\ce{H2} systems were obtained from a direct implementation of the FCI exchange-dispersion energy based on transition density matrices from full  response equations (see Supporting Information for details). The results have been tested against a different, house-developed code specific for interactions between two two-electron monomers in singlet states.
In this case solutions corresponding to the considered physical singlet state of the dimer may be accessed by projection onto the appropriate irreducible representation of the symmetric S$_4$ group (represented by the Young tableau [2$^2$]).~\cite{Korona:97}

Apart from calculations for model few-electron system, we verified the accuracy of the ERPA-based exchange-dispersion energies for many-electron dimers from the data set introduced by Korona\cite{Korona:13} (denoted TK21) and dimers from the A24 data set\cite{Rezac:13} by \v{R}ez\'{a}\v{c} and Hobza. The same qualitative trends were observed for both sets. In Section~\ref{sec:sref} we present TK21 results in the aug-cc-pVDZ basis set for which benchmark SAPT(CCSD) values are available.~\cite{Korona:13} The aug-cc-pVTZ results for both TK21 and A24 data sets are provided in the Supporting Information.

When the reference $E_{\rm exch-disp,ref}$ energy is known, relative percent errors are obtained with the formula
\begin{equation}
\Delta = \frac{ \, E_{\rm exch-disp}-E_{\rm exch-disp,ref}}{|E_{{\rm exch-disp,ref}}|} \cdot 100\% \ \ \ .
\label{eq:delta}
\end{equation}
The negative and positive values of $\Delta$ correspond to under- and overestimation of the magnitude of the reference exchange-dispersion energy. 

\section{Results \label{sec:res}}

\subsection{Multireference model systems: H$_2$$\cdots$H$_2$, Be$\cdots$Be and He$\cdots$H$_2$ \label{sec:res:model}}

In this section we examine the accuracy of the ERPA-based exchange-dispersion energy in three model systems of a multireference character: \ce{H2}$\cdots$\ce{H2}, Be$\cdots$Be and He$\cdots$\ce{H2}(1 $^1\Sigma_g^+$, 1 $^1\Sigma_u^+$, 1 $^1\Pi_u$) dimers. In each of the monomers (He, \ce{H2} and Be) two electrons were kept active.

\begin{figure*}
\centering
\includegraphics[width=\textwidth]{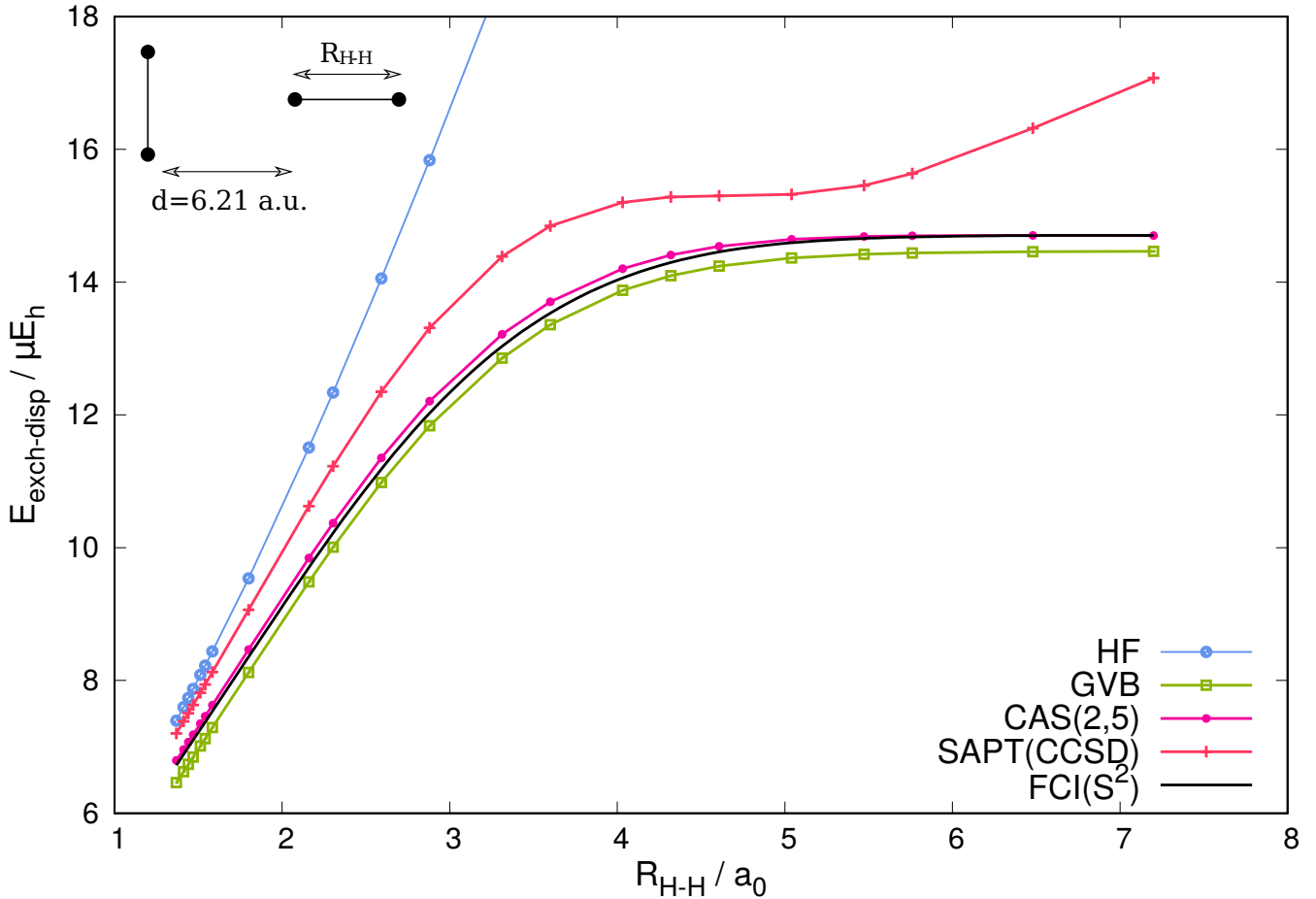}
\caption{Exchange-dispersion energy at the Hartree-Fock (HF), GVB, CASSCF, CCSD and FCI($S^2$) levels of theory for the \ce{H2}$\cdots$\ce{H2} dimer in the T-shaped configuration. The intermolecular distance is fixed at 6.21$\,a_0$. In one of the monomers the $R_{\rm H-H}$ distance is varied from 1.37 to 9$\,a_0$ while in the other monomer the H--H bond is kept at a fixed distance of 1.44$\,a_0$. CAS(2,5) refers to the active space of each monomer.}
\label{fig:1}
\end{figure*}

We begin with the analysis of the \ce{H2}$\cdots$\ce{H2} system. The dimer is kept in the T-shaped geometry and the multireference character is increased by elongation of the covalent bond in one of the interacting monomers (see Figure~\ref{fig:1} and Ref.~\onlinecite{Hapka:19} for details). 
Accurate description of the interaction energy along the dissociation coordinate has been shown to pose enormous challenge not only for single-reference methods but also methods providing multireference description of monomers \cite{Hapka:19,Brzek:19}. Interaction between two hydrogen molecules, one of them undergoing dissociation, emerges from interplay of the long-range dynamic and nondynamic correlation effects. Thus, it is a perfect model system for our approach.
In Figure~\ref{fig:1} we compare results obtained with single-reference (Hartree-Fock and CCSD) and multireference (CAS and GVB) wave functions in the aug-cc-pVTZ basis set.

The exchange-dispersion energy calculated at the Hartree-Fock level of theory is overestimated by approximately $16\%$ already in the vicinity of the minimum ($R_{\rm H-H} = 1.44~a_0$) compared to the FCI benchmark. As expected, the error rises rapidly when the H-H bond is stretched. Somewhat surprisingly, SAPT(CCSD) also overestimates by as much as $7\%$ near the equilibrium distance. 
Error of this magnitude in the minimum region, together with the qualitatively wrong behavior observed in the $R_{\rm H-H} > 4.5~a_0$ regime, should be attributed to two factors: the neglect of the cumulant part in the exchange-dispersion expression, as well as approximations in the time-independent CCSD propagators (in our calculations the CCSD(3)\cite{Korona:06} model was applied).

In contrast to the single-reference case, the multireference description of the monomers within the ERPA framework provides a quantitative agreement with the FCI benchmark. With the GVB ansatz the relative percent errors do not exceed $-4.0\%$ for the entire curve, which corresponds to slight underestimation of the exchange-dispersion energy. The choice of the CAS(2,5) wave function for each hydrogen molecule (notice that choosing only two active orbitals in CAS, i.e. the (2,2) active space, would make CAS wave function identical to that of GVB) reduces the error to below $1.5\%$. This level of accuracy matches the one reported for the dispersion component.~\cite{Hapka:19}

In Table~\ref{tab:be} we present dispersion and exchange-dispersion energies obtained for the beryllium dimer near the equilibrium geometry ($4.7~a_0$). Our results show that the essential part of both energy components is captured already by the CAS(2,5) wave function for each Be atom. Extension of the active space to 14 orbitals brings only a minor change of ca.~$0.5\%$. As discussed in Ref~\onlinecite{Hapka:19}, the dispersion interaction in the beryllium dimer obtained at the ERPA-CAS level of theory remains in excellent agreement with the SAPT(CCSD) result. This no longer holds for exchange-dispersion: while the SAPT(CCSD) value is 3.327~m$E_{\rm h}$, our result based on the CAS(2,14) reference amounts to 2.651~m$E_{\rm h}$. 

The single-reference-based SAPT(CCSD) value cannot be considered a benchmark for the beryllium dimer. Taking into account that the ERPA approximation yields accurate one-electron response function for beryllium,~\cite{Pernal:12} we expect that multireference ERPA-CAS calculations provide a more reliable estimate of the exchange-dispersion energy in this system. The remaining source of error in our approach is the quality of the ERPA-based two-electron transition density matrices. Since each Be atom is effectively treated as a two-electron system (core electrons are kept frozen), we expect that 2-TRDMs are described with satisfactory accuracy, similar to the \ce{H2}$\cdots$\ce{H2} case. 

In Table~\ref{tab:be} we also report the dispersion and exchange-dispersion energy values for beryllium dimer in the uncoupled approximation. Unfortunately, this approximation leads to significant errors underestimating the dispersion contributions by $-21\%$ and exchange-dispersion by $-34\%$ with respect to coupled results. 

With the HF description of the monomers, for the exchange-dispersion energy one obtains the value as high as 4.671~m$E_{\rm h}$ thus deviating largely from both CAS and SAPT(CCSD) values. 
The observed discrepancy reflects the influence of static correlation effects, which cannot be correctly accounted for in single-reference calculations. It should finally be noted that the HF exchange-dispersion energy following from the formula proposed by Hesselmann \textit{et al.}\cite{Hesselmann:05} and implemented in the \textsc{Molpro} package\cite{Molpro:12} amounts to 2.237~m$E_{\rm h}$ which accidentally remains in better agreement with the ERPA-based result.

\begin{table*}
\centering
\caption{Dispersion and exchange-dispersion energies (in m$E_{\rm h}$) at the coupled and uncoupled (UC) levels of theory for the Be dimer at 4.7$\,a_0$ separation. All calculations were performed in the aug-cc-pVTZ basis set. }
\begin{tabular}{l S S p{2mm} S S}
\hline
\multicolumn{1}{l}{Method} & \multicolumn{1}{c}{$E_{\rm disp}$} & \multicolumn{1}{c}{$E^{\rm UC}_{\rm disp}$} && \multicolumn{1}{c}{$E_{\rm exch-disp}$} & \multicolumn{1}{c}{$E^{\rm UC}_{\rm exch-disp}$} \\ \hline
CAS(2,5)   & -18.52  & -12.64            && 2.636 & 1.789 \\
CAS(2,14)  & -18.61  & -14.77            && 2.651 & 1.745 \\
HF         & -20.14  & -18.30            && 4.671 & 1.626 \\
SAPT(CCSD) & -18.86  &                   && 3.327 &  \\ \hline
\end{tabular}
\label{tab:be}
\end{table*}

\begin{figure*}
\centering	
\includegraphics[width=\textwidth]{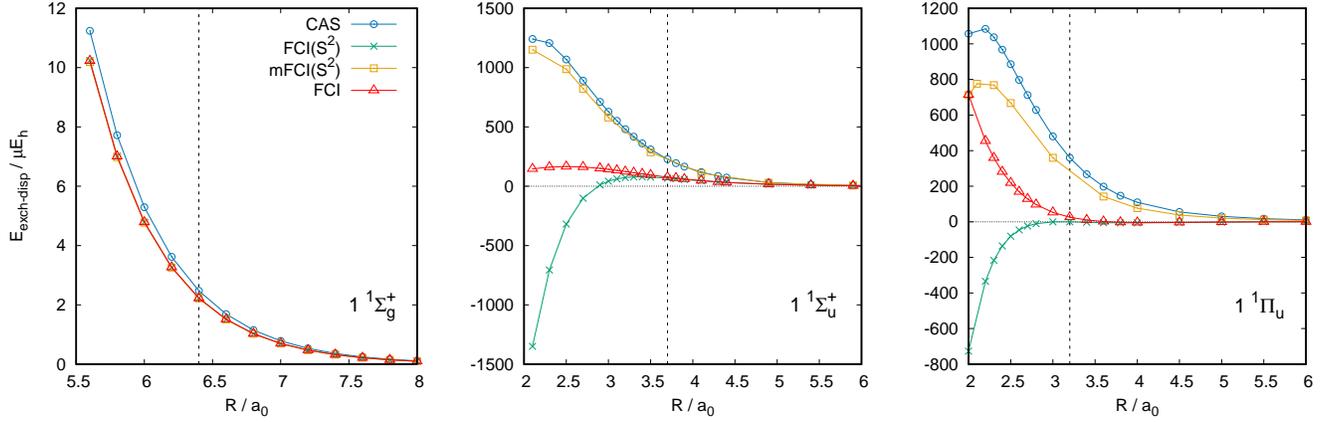}
\caption{Exchange-dispersion energy (in $\mu E_h$) at the FCI, FCI($S^2$) and CASSCF levels of theory for the \ce{He}$\cdots$\ce{H2}(1 $^1\Sigma_g^+$, 1 $^1\Sigma_u^+$, 1 $^1\Pi_u$) dimer in the T-shaped configuration. Basis set used is d-aug-cc-pVTZ. $R$ denotes the distance between the helium atom and the center of mass of \ce{H2}. Vertical dotted lines mark positions of the van der Waals minima. The He atom and the \ce{H2} molecule are described with CAS(2,2) and CAS(2,8) wave functions, respectively.}
\label{fig:2}
\end{figure*}

The proposed formalism is valid not only for ground states, but also when one or both monomers are in excited states, on condition that  a zeroth-order wavefunction of a dimer is not degenerate. To investigate the performance of the ERPA-based approximation to the exchange-dispersion energy for excited states, we examine the \ce{He}$\cdots$\ce{H2}($X$) complex, where
$X$ denotes either the ground state ($^1\Sigma_g^+$) or one of the two lowest singlet excited states ($^1\Sigma_u^+$ and $^1\Pi_u$) of the hydrogen molecule. The dimer is kept in the T-shaped configuration in which we vary the distance between He and the center of mass of \ce{H2}. The helium atom was described with a CAS(2,2) reference, whereas for the \ce{H2} molecule a CAS(2,8) wavefunction was used.
In the case of the \ce{He}$\cdots$\ce{H2}($^1\Pi_u$) dimer we report results for the $A''$ state, which corresponds to perpendicular orientation of the $\pi$ orbital of \ce{H2} with respect to the plane of the complex. The bond lengths assumed for the hydrogen molecule correspond to the equilibrium values for the investigated states and they read 1.401$\,a_0$, 2.443$\,a_0$, and 1.951$\,a_0$, for $^1\Sigma_g^+$, $^1\Sigma_u^+$, and $^1\Pi_u$ states, respectively. Figure~\ref{fig:2} presents the results for the exchange-dispersion energy computed either exactly (``FCI'' curves), i.e., by describing monomers with FCI wave functions and employing the expression shown in Eq.~\eqref{eq:exdF}, or approximately (``CAS'' curves), which consists of using CASSCF for monomers combined with the proposed ERPA-based approximation for the exchange-dispersion energy, Eqs.~\eqref{eq:exdmat}-\eqref{eq:intW}. In addition, we present FCI results in the $S^2$ approximation (``FCI($S^2$)'' curves, see also Supporting Information for details).

First, let us compare the FCI and FCI($S^2$) results. In general, the single-exchange approximation is expected to fail at distances shorter than the van der Waals minimum.~\cite{Schaffer:12,Schaffer:13} As could be inferred from Figure~\ref{fig:2}, the breakdown of the $S^2$ approximation is more rapid for both excited \ce{He}$\cdots$\ce{H2}($^1\Sigma_u^+$, $^1\Pi_u$) states of the complex than for the ground \ce{He}$\cdots$\ce{H2}($^1\Sigma_g^+$) state. This is not surprising---the minimum in the ground-state occurs at $6.4\,a_0$, whereas in excited states the interaction is considerably stronger and the minima shift to ca. $3.7\,a_0$ and $3.2\,a_0$ for the $^1\Sigma_u^+$ and $^1\Pi_u$ and states, respectively.

Next, we assess the performance of the ERPA-based approach. For the single-reference dominated \ce{He}$\cdots$\ce{H2}($^1\Sigma_g^+$) state ERPA combined with CAS remains in good agreement with the FCI benchmark (Figure~\ref{fig:2}), with relative percent errors remaining in the $3-6\%$ range for the entire curve. 
In contrast, for both excited states of the dimer our method fails qualitatively. The discrepancy with respect to the benchmark result is huge not only around van der Waals minima, but also at larger distances which shows that the single-exchange approximation is not the main source of error in this case.

We have verified that the erroneous behavior observed for excited states of \ce{H2} can be traced to the fact that ERPA does not recover double excitations in the orbital active space.~\cite{Pernal:12,Pernal:18}
For \ce{H2} these excitations can only be described if the diagonal elements of the response eigenvectors\cite{Giesbertz:12} are retained (see Supporting Information for details). Notice that in the approximations assumed to obtain the exchange-dispersion energy we only recover off-diagonal elements of the transition density matrices, cf.\ Eq.(\ref{eq:1trdm}). Consequently, the response function and response properties in dispersion components of the interaction energy are impaired, since double excitations are thrown out.

To illustrate the influence of double excitations on the quality of exchange-dispersion in this system, we discard them in the FCI linear response, meaning in practice that we neglect the diagonal elements of the 1-TRDM's. The resulting exchange-dispersion energy is plotted in the S$^2$ approximation in Figure~\ref{fig:2} (``mFCI(S$^2$)'' curves). For dimers in excited states the mFCI(S$^2$) curves are dramatically different compared to the FCI(S$^2$) ones and they resemble ERPA-based CAS results. 
The lack of double excitations in ERPA has much lesser effect on the $E^{(2)}_{\rm disp}$ component---errors with respect to FCI remain in the $8-30$\% range for the He$\cdots$\ce{H2}($^1\Sigma_u^+$) state of the dimer, and in the $7-12$\% range for the He$\cdots$\ce{H2}($^1\Pi_u$) state (see Figure~S1 and Table~S2 in the Supporting Information). Taking into account that $E^{(2)}_{\rm exch-disp}$ is small in comparison to $E^{(2)}_{\rm disp}$, and also the fact that the overall contribution from double excitations is expected to be less pronounced for many-electron systems than for the hydrogen molecule,~\cite{Pernal:14b} one can still expect that the proposed ERPA-based approximations will prove useful when applied to dispersion interactions in excited systems.

\subsection{Single-reference systems \label{sec:sref}}

In order to verify the accuracy of the exchange-dispersion energies based on the ERPA response for many electron monomers, we analyze dimers from the TK21 data set.~\cite{Korona:13} All of them are single-reference systems, i.e., the pertinent multiconfiguration wave functions are dominated by a single determinant.
Our results based on GVB and CASSCF description of the monomers are compared with the SAPT(CCSD) benchmark.

\begin{table*}
\centering
\caption{Summary of error statistics (in percent) for the exchange-dispersion energy calculated within coupled ($E_{\rm exch-disp}$) and uncoupled ($E^{\rm UC}_{\rm exch-disp}$) approximations with Hartree-Fock (HF), GVB and CASSCF references for dimers of the TK21 data set. Errors are given with respect to the SAPT(CCSD) values.}
\begin{tabular}{l p{3mm} S S S p{4mm} S S S}
\hline
                              && \multicolumn{3}{c}{$E_{\rm exch-disp}$} && \multicolumn{3}{c}{$E^{\rm UC}_{\rm exch-disp}$} \\ 
                              && HF  & GVB & CAS && HF  & GVB & CAS \\
                              \cline{3-5} \cline{7-9}
$\overline{\Delta}$           && -12.67 & -18.66 & -12.02 && -16.49 & -35.73 & -43.19 \\
$\sigma$                      &&  14.48 &  10.22 &  7.29  &&  16.54 &  12.87 &  23.65 \\
$\overline{\Delta}_{\rm abs}$ &&  17.68 &  18.66 &  12.22 &&  21.78 &  35.73 &  43.19 \\
$\Delta_{\rm max}$            &&  27.42 &  38.24 &  25.64 &&  33.33 &  61.56 &  87.93 \\
\hline
\end{tabular}
\label{tab:tk21}
\end{table*}

\begin{figure*}
\includegraphics[width=\textwidth]{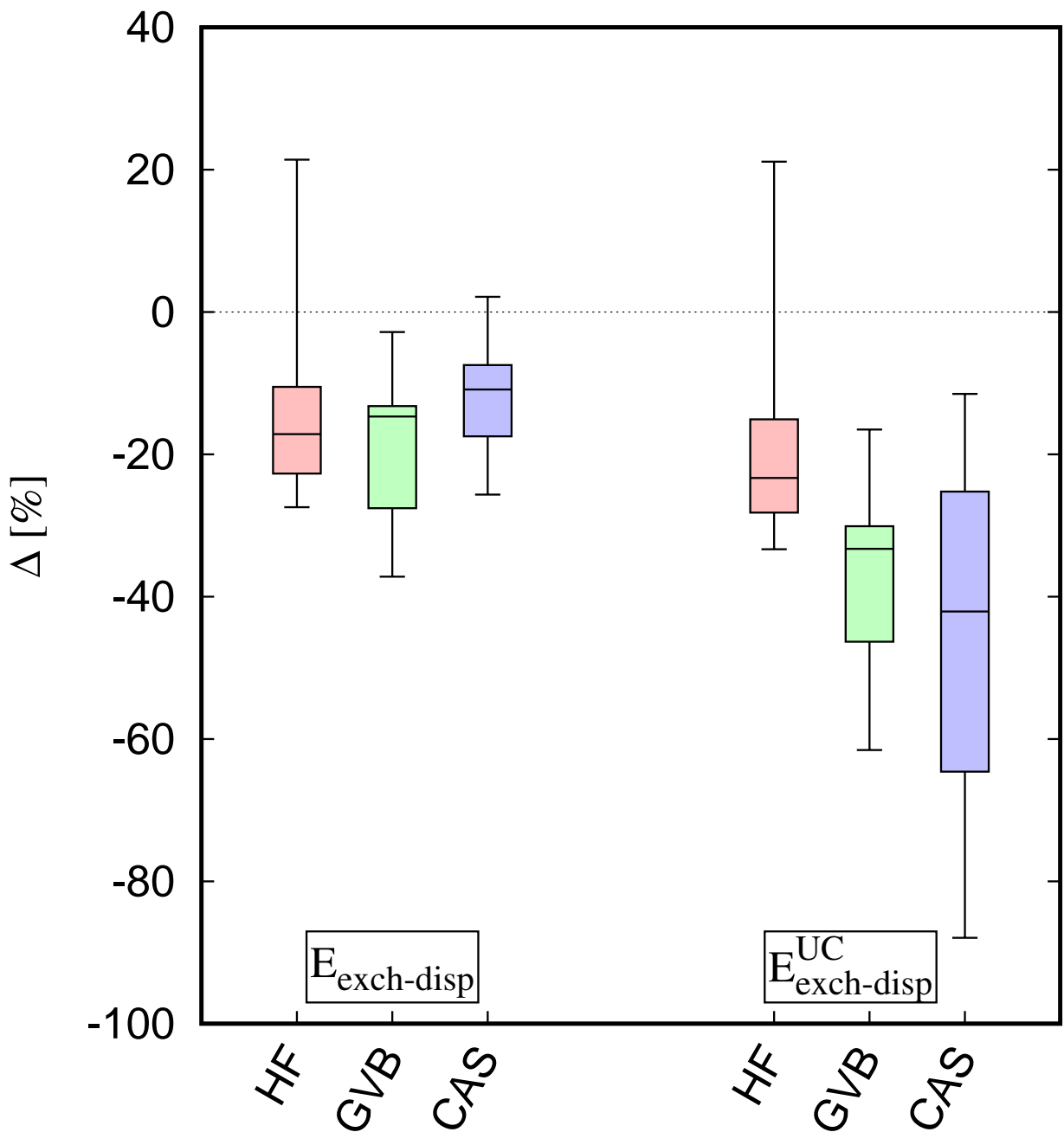}
\caption{Box plots of relative percent errors in the calculated exchange-dispersion energies in the coupled (E$_{\rm exch-disp}$) and uncoupled (E$^{\rm UC}_{\rm exch-disp}$) approximations for dimers of the TK21 data set. HF, GVB and CAS denote wavefunction description of the monomers. Errors are given with respect to the SAPT(CCSD) reference. The box and outer fences encompass 50\% and 100\% of the distribution, respectively.}
\label{fig:tk21}
\end{figure*}

The relative percent error $\Delta$ for each system is calculated according to Eq.~\eqref{eq:delta}. In Table~\ref{tab:tk21} we show the error statistics for the TK21 data set using mean error $\overline{\Delta}$, the standard deviation $\sigma$, the mean absolute error $\overline{\Delta}_{\rm abs}$, and the maximum absolute error $\Delta_{\rm max}$ as error measures. The errors are also presented in terms of box plots in Figure~\ref{fig:tk21}. In both Table~\ref{tab:tk21} and Figure~\ref{fig:tk21} the Hartree-Fock data are given for comparison.

As it is known, CASSCF wavefunctions are constructed from a small number of active orbitals only to capture major static correlation effects but they lack dynamic correlation. The lack of intramonomer dynamic correlation in all of the analyzed methods results in substantial underestimation of the exchange-dispersion energy with absolute errors exceeding $12\%$ (Table~\ref{tab:tk21}).
Treatment of the monomers at the CASSCF level ($\overline{\Delta} = -12.0\%$, $\overline{\Delta}_{\rm abs} =12.2\%$) offers a modest improvement over the Hartree-Fock wavefunction ($\overline{\Delta} = -12.7\%$, $\overline{\Delta}_{\rm abs} = 17.7\%$). Worth noting, however, is a nearly twofold reduction in the standard deviation: from $\sigma = 14.5\%$ in HF to $\sigma = 7.3\%$ in CASSCF.  GVB is inferior to both CAS- and HF-based variants in terms of mean errors ($\overline{\Delta} = -\overline{\Delta}_{\rm abs} = -18.7\%$), although the spread of errors ($\sigma = 10.2\%$) is smaller than in the Hartree-Fock case. 

As evident from both Table~\ref{tab:tk21} and Figure~\ref{fig:tk21}, the uncoupled approximation to the exchange-dispersion energy leads to prohibitively large errors in the ERPA-based approaches. This remains in agreement with our previous results for the dispersion components.~\cite{Hapka:19} A particularly poor performance with errors above $80\%$ is observed for the \ce{CH4}-\ce{CH4} and \ce{P2}-\ce{P2} dimers (also for methane-containing complexes in the A24 data set, see Table S7). In contrast, at the HF level of theory the quality of the uncoupled exchange-dispersion energy is only slightly worse than in the coupled formulation.

Finally, we observe that the errors in exchange-dispersion energies reported in Table~\ref{tab:tk21} and Figure~\ref{fig:tk21}
are overall higher than those of the second-order dispersion energy presented in Ref.~\onlinecite{Hapka:19}. For instance, the mean absolute error for the TK21 data set in dispersion calculated using CASSCF description of monomers amounts to $7.0\%$ (cf.\ Table~S5 in Ref.~\onlinecite{Hapka:19}) compared to $12.2\%$ error in exchange-dispersion (Table~\ref{tab:tk21}). 
Fortunately, it is expected that some cancellation of error will occur, since exchange-dispersion dampens only ca.\ $5-15\%$ of the dispersion energy in the van der Waals minimum, the components are of the opposite signs, and both of them are underestimated with respect to the CCSD benchmark. 
In the aforementioned case the error of the total $E^{(2)}_{\rm DISP} = E^{(2)}_{\rm disp} + E^{(2)}_{\rm exch-disp}$ term amounts to $5.9\%$ (see Table S8 in the Supplementary Information).

The accuracy of second-order dispersion energy components may suffer from both the lack of intramonomer dynamic correlation in monomer wave functions as well as from approximate treatment of one-electron TRDMs. Compared to the dispersion contribution, the exchange-dispersion energy expression requires access to two-electron TRDMs, which in our method are described within the REOM-ERPA formalism. This additional approximation is the most likely cause of larger errors for the exchange-dispersion energies than for the dispersion ones.

\section{Conclusions \label{sec:concl}}

We have presented a general formulation of the second-order exchange-dispersion SAPT energy in the single-exchange approximation applicable to dimers, in which at least one of the interacting monomers requires a description by a multireference wave function. 
The derived formula is based on employing the Rowe's equation of motion formalism in the extended random phase approximation
to obtain one- and two-electron response properties. The final formula defined by Eq.~\eqref{eq:exdmat}-\eqref{eq:intW} requires access to only 1- and 2-TRDMs of the monomers. It is valid for both ground and nondegenerate excited states of the monomers in spin singlet states. The proposed approach was applied in combination with either CASSCF or GVB treatment of the interacting systems. This work complements our recent study\cite{Hapka:19} on the ERPA-based variant of the second-order dispersion energy.  

We have analyzed the accuracy of our multireference approach for several model, few-electron systems. For the \ce{H2}$\cdots$\ce{H2} dimer, in which static correlation effects are introduced by stretching one of the H-H bonds, the ERPA-based exchange-dispersion energy remains in excellent agreement with the FCI benchmark. As expected, results from single-reference SAPT variants deviate from the FCI reference in the strongly-correlated regime. 
In a similar way, for the challenging case of the beryllium dimer\cite{Patkowski:07} our approach based on CAS(2,14) wave functions gives exchange-dispersion energy
which is substantially different from both the HF-SAPT and SAPT(CCSD) values.

Results for the lowest excited states of the \ce{He}$\cdots$\ce{H2} dimer (excitation localized on the hydrogen molecule) demonstrate that the exchange-dispersion energy based on REOM-ERPA approximation may be qualitatively wrong due to the lack of double diagonal excitations. Nevertheless, the total dispersion interaction, i.e.\ the sum of the second-order dispersion and exchange-dispersion components, remains qualitatively correct, as the dominating polarization $E^{(2)}_{\rm disp}$ contribution is much less affected by the missing excitations. 

Excellent quality of exchange-dispersion energies obtained with our approach for ground-state, few-electron dimers illustrates that it is possible to recover intramonomer correlation effects in these molecules using either GVB reference or CAS wave functions with small active spaces. 
However, for many-electron systems a substantial amount of intramonomer correlation is missing even if the monomers are described at the CAS($n,n$) ($n$ active electrons on $n$ active orbitals) level of theory. This results in sizeable errors in exchange-dispersion energies. For dimers of the TK21 and A24 data sets, which do not include strongly correlated electrons, ERPA combined with CASSCF affords exchange-dispersion energies only slightly more accurate than the Hartree-Fock-based SAPT. The GVB-based results are slightly inferior to both CASSCF and Hartree-Fock. Similar behavior was observed for the $E^{(2)}_{\rm disp}$ energy in our previous work.~\cite{Hapka:19} It should be emphasized, that the true target of the ERPA-based SAPT variant, are many-electron systems of a multirefence character for which neither Hartree-Fock nor other single-reference SAPT formulations are reliable.

Evaluation of the density-matrix-based exchange-dispersion formula in the ERPA approximation scales with the sixth power in terms of the orbital basis size. 
The overall cost of the calculation is dominated by steps scaling as $n_{\rm OCC}^3 n_{\rm SEC}^3$. The computational burden may be significantly reduced by invoking the uncoupled approximation which circumvents diagonalization of the full ERPA eigenproblem. Unfortunately, results for model systems investigated both in this and in our previous work indicate that the uncoupled approximation should be avoided in the EPRA-based SAPT formulation, as it leads to prohibitevely large errors. 
Thus, further development of the proposed scheme aiming at computation cost reduction should employ density fitting or development of the local variant.

\begin{acknowledgement}
The authors would like to thank Grzegorz Cha{\l}asi{\'n}ski for helpful discussions and commenting on the manuscript.

K. P. and M. H. were supported by the National Science Centre of Poland under Grant No. 2016/23/B/ST4/02848.
\end{acknowledgement}

\newpage
\providecommand{\latin}[1]{#1}
\makeatletter
\providecommand{\doi}
  {\begingroup\let\do\@makeother\dospecials
  \catcode`\{=1 \catcode`\}=2 \doi@aux}
\providecommand{\doi@aux}[1]{\endgroup\texttt{#1}}
\makeatother
\providecommand*\mcitethebibliography{\thebibliography}
\csname @ifundefined\endcsname{endmcitethebibliography}
  {\let\endmcitethebibliography\endthebibliography}{}

\end{document}